\documentclass[acmsmall,screen]{acmart}
\AtBeginDocument{%
  }

\setcopyright{acmlicensed}
\copyrightyear{2024}
\acmYear{2024}
\acmDOI{XXXXXXX.XXXXXXX}

\usepackage{graphicx} 
\usepackage{xspace}
\usepackage{enumitem} 
\usepackage{multirow}
\usepackage{multicol}
\usepackage{ulem}
\usepackage{tcolorbox}
\usepackage{makecell}
\usepackage{pifont}
\usepackage{soul}
\usepackage{hyperref}
\usepackage{multirow}
\usepackage{listings}
\usepackage{fancybox}
\usepackage{enumitem}
\usepackage{graphicx}
\usepackage{color}
\usepackage{array}
\usepackage{amsmath}
\usepackage{xspace}
\usepackage{url}
\usepackage{tikz}
\usepackage{caption}
\usepackage{pgfplots}
\usepackage{balance}
\usepackage{subcaption}
\pgfplotsset{compat=1.16}
\usepackage{pgf-pie}
\usetikzlibrary{pgfplots.statistics,calc}
\usepackage{algorithm}
\usepackage{algorithmic}
\usepackage{adjustbox}
\usepackage{makecell}

\usepackage{tcolorbox}
\makeatletter
\newcommand{\mybox}[1]{%
	\setbox0=\hbox{#1}%
	\setlength{\@tempdima}{\dimexpr\wd0+13pt}%
	\begin{tcolorbox}[boxrule=0.5pt, colback=white, arc=4pt,
		left=6pt,right=6pt,top=6pt,bottom=6pt,boxsep=0pt]
		#1
	\end{tcolorbox}
}

\definecolor{songcolor}{RGB}{191,191,191}

\newcommand{\q}[1]{``{#1}''}
\newcommand{\code}[1]{\texttt{#1}\xspace}
\newcommand{\tool}{TITAN\xspace}

\title{Task-oriented Prompt Enhancement via Script Generation}
\author{Chung-Yu Wang}
\affiliation{
  \institution{York University}
  \city{Toronto}
  \country{Canada}
}
\email{cywang14@yorku.ca}

\author{Alireza DaghighFarsoodeh}
\affiliation{%
  \institution{York University}
  \city{Toronto}
  \country{Canada}
}
\email{aliredaq@yorku.ca}

\author{Hung Viet Pham}
\affiliation{%
  \institution{York University}
  \city{Toronto}
  \country{Canada}
}
\email{hvpham@yorku.ca}

\date{June 2024}

\begin{document}

\begin{abstract}
    Large language Models (LLMs) have demonstrated remarkable abilities across various tasks, leveraging advanced reasoning. Yet, they struggle with task-oriented prompts due to a lack of specific prior knowledge of the task answers. The current state-of-the-art approach, PAL, utilizes code generation to address this issue. 
    However, PAL
    depends on manually crafted prompt templates and examples while still producing 
    inaccurate results.
    
    In this work, we present \tool---a novel strategy designed to enhance LLMs’ performance on task-oriented prompts. \tool achieves this by generating scripts using a universal approach and zero-shot learning. Unlike existing methods, \tool eliminates the need for detailed task-specific instructions and extensive manual efforts. \tool enhances LLMs' performance on various tasks by utilizing their analytical and code-generation capabilities in a streamlined process.
    \tool employs two key techniques: (1) step-back prompting to extract the task's input specifications and (2) chain-of-thought prompting to identify required procedural steps. This information is used to improve the LLMs' code-generation process. \tool further refines the generated script through post-processing and the script is executed to retrieve the final answer.
    
    Our comprehensive evaluation demonstrates \tool's effectiveness in a diverse set of tasks.
    On average, \tool outperforms the state-of-the-art zero-shot approach by 7.6\% and 3.9\% when paired with GPT-3.5 and GPT-4.
    Overall, without human annotation, \tool achieves state-of-the-art performance in 8 out of 11 cases
    while only marginally losing to few-shot approaches (which needed human intervention) on three occasions 
    by small margins.
    This work represents a significant advancement in addressing task-oriented prompts,
    offering a novel solution for effectively utilizing LLMs in 
    everyday life tasks.
\end{abstract}



\maketitle
\section{Introduction}

Large Language Models (LLMs) like GPT have significantly advanced the field of Natural Language Processing (NLP) through their proficiency in a wide range of tasks, including text generation~\cite{lu2023bounding,abdullin2024synthetic}, translation~\cite{he2024exploring,wang2023document}, summarization~\cite{jin2024comprehensive}, and answering questions~\cite{kim2023sure,li2024flexkbqa}. These models are trained on vast datasets, enabling them to produce text that is both coherent and contextually appropriate~\cite{wu2023survey}. However, when it comes to handling basic, task-oriented problems that involve numerical calculations or step executions, LLMs often fall short~\cite{yu2023metamath, li2023metaagents}.
This is an inherent weakness due to the way LLMs are designed and constructed.
LLMs are trained on large text corpora and finely tuned for linguistic content creation and processing~\cite{ahn2024large}.
LLMs construct answers from text corpora and often face difficulties when the answers are not present directly in the training dataset but require 
precise numerical operations or step executions~\cite{goertzel2023generative}.
For instance, LLMs' limitation in counting tasks is evident in a simple test. If GPT-4~\cite{achiam2023gpt} using greedy decoding (0.0 temperature), is asked: \q{Ed had 22 more marbles than Doug. Doug lost 8  of his marbles at the playground. How many more marbles did Ed have than Doug then?}, the answer would be: \q{Ed still had 22 more marbles than Doug}.
LLMs have 
a tendency to provide approximations or incorrect counts.
This highlights the necessity for specialized prompt improvement
to address these kinds of task-oriented 
challenges.

One approach is to
utilize prompt engineering \cite{zheng2023take,liu2022generated,yao2024tree,madaan2024self,zhou2023leasttomost} to improve LLMs' performance on specific tasks. This process involves crafting well-defined, strategically structured prompts to guide models towards specific outcomes. For instance, Chain-of-Thought (CoT)~\cite{wei2022chain}, one of the prompting techniques, breaks down problems into intermediate, textual steps to facilitate problem understanding and reveal steps toward potential solutions. 
CoT could be applied to the above problems to help LLM derive steps to complete the task. However, CoT still struggles with task-oriented problems as it still inherits the LLMs' weakness of not having direct access to the numerical answers~\cite{goertzel2023generative}. To address this, prior work proposes Program-Aided Language Models (PAL)\cite{gao2023pal}. PAL employs the generation of code as an intermediate step in the reasoning process to bridge the task execution gap.

Nevertheless, these prompt techniques perform optimally when used with few-shot prompting, in which hand-picked problem examples and answers are provided to help LLMs correctly comprehend and solve problems. However, crafting these examples for few-shot prompting is non-trivial and is an extra burden for the users without any related experience~\cite{dang2022prompt,liu2021makes,perez2021true}. Using wrong examples~\cite{rubin2021learning} or wrongly ordering examples~\cite{lu2021fantastically} for few-shot prompting will affect performance dramatically. To address the drawbacks of crafting few-shot prompts for task-oriented problems, we introduce \tool, a novel prompting framework to address the challenges of task-oriented problems while requiring no user effort.



Distinct from PAL, 
which directly generates scripts, 
\tool incorporates two additional intermediate reasoning stages: input extraction and step extraction. These additional reasoning stages facilitate the generation of accurate scripts without the need for labeled examples.
Specifically, \tool applies step-back prompting~\cite{zheng2023take} to extract the inputs and their specifications for each task, making it the first framework to incorporate step-back prompting into code generation tasks. Input extraction helps LLMs overcome their tendency to perform poorly when given meaningless variables or when there is a misunderstanding of the problem's inputs.
In parallel, \tool extracts procedure steps to complete the task by utilizing CoT prompting, which has been shown to aid LLMs in accurately comprehending problems by breaking them down into steps~\cite{wei2022chain,kojima2022large}.
Finally, \tool combines the information from these two additional reasoning stages to generate the most accurate scripts that are capable of producing precise answers. 

To assess the effectiveness of \tool, we evaluate it
across seven distinct prior datasets~\cite{cobbe2021training,gao2023pal,patel2021nlp,miao2021diverse,koncel2016mawps,suzgun2022challenging} containing
mathematical and symbolic reasoning tasks and four additional task-oriented benchmarks we constructed in this work.
The results demonstrate that \tool, when paired with GPT-4, consistently outperforms PAL zero-shot variant
by an average accuracy improvement of 3.9\%. When compared to few-shot approaches, \tool performs better or comparable in most cases (8 out of 11 datasets).
Furthermore, 
\tool remains effective even with less advanced LLMs, such as GPT-3.5. \tool improves state-of-the-art zero-shot approaches on GPT-3.5 by an average of 7.6\%.
Overall, when using GPT-4, \tool achieves state-of-the-art performance on 8 out of 11 evaluated datasets while only marginally losing to few-shot approaches (which needed human intervention) on three datasets by small margins.
Our ablation study further indicates that
the addition of input and step extraction stages enables \tool to tackle task-oriented problems
accurately without the need for hand-crafted
examples and prompt templates.

This work's main contributions are as follows:
\begin{itemize}[leftmargin=*,topsep=0pt]
    \item A new \tool framework that can improve LLMs' ability to solve task-oriented problems in a zero-shot manner by extracting inputs and procedure steps using step-back and chain-of-thought prompting respectively.
    \item Four task-oriented datasets consist of Finding, Counting, True/False, and Generative problems that can be used to better assess the capacity of LLMs in addressing task-oriented challenges.
    \item An extensive evaluation of \tool on eleven datasets employing two different popular versions of LLMs (GPT-3.5 Turbo and GPT-4).
\end{itemize}



\section{Approach}
\begin{figure}
    \includegraphics[width=\linewidth]{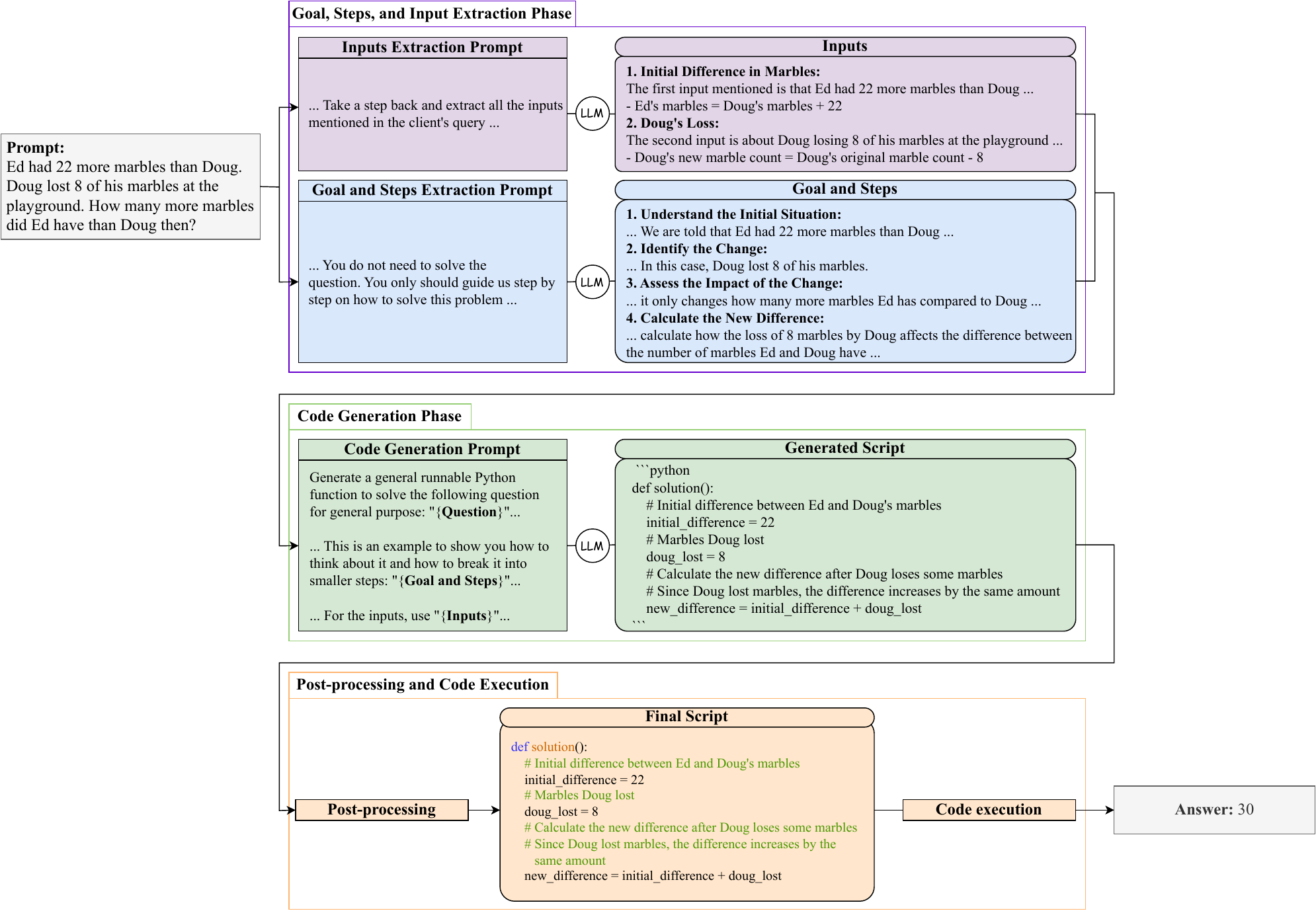}
    \caption{\tool Overview}  
    \label{fig:Example}
\end{figure}
In this work, we introduce \tool, a novel approach to improve LLMs responses to task-oriented prompts.
This is achieved by generating code through the use of a universal prompt template coupled with zero-shot learning techniques. 
\tool significantly differs from prior prompt-specific strategies~\cite{gao2023pal, chen2023program,zhou2023solving,zhao2023automatic,liu2023plan,zhang2024self,chen2024boosting} since
it provides a general and effective way to boost the performance of LLMs in performing various tasks without the need for individual prompt adjustments.
\tool enhances LLMs in two key ways: firstly, by utilizing the inherent abilities of LLMs to dissect and analyze complex queries; and secondly, by employing the code generation ability of LLMs to create executable scripts that produce the answers.


\subsection{Challenges}
As discussed in the Introduction, task-oriented prompts that involve numerical numbers (e.g., Ed had 22 more marbles than Doug. Doug lost 8 of his marbles at the playground. How many more marbles did Ed have than Doug then?) are not a direct fit for LLMs~\cite{goertzel2023generative}. LLMs excel at analyzing and understanding complex queries by learning prior knowledge from a large corpus of text data~\cite{chang2023survey,wang2023empower}. However, for LLMs to correctly respond to a query, the answer must be composed of such prior knowledge~\cite{ge2024openagi}. In the case of task-oriented prompts, the answer often is the result of an execution of some procedure described by the prompt. Hence in most cases, the result does not directly exist in the prior knowledge.

Conversely, LLMs have shown proficiency in generating code from natural language descriptions~\cite{dehaerne2022code, shin2021survey, svyatkovskiy2020intellicode, yeticstiren2023evaluating, liu2023improving}, as their training datasets contain numerous coding instances that the LLMs can utilize. By combining the analysis ability of LLMs with the task execution function of computer programs, we can create a system that is capable of precisely performing complex task-oriented prompts.

Prior approaches improve task-oriented prompts with code generation by relying heavily on task-specific prompts, requiring extensive manual effort and domain expertise. Such approaches~\cite{gao2023pal, chen2023program, zhao2023automatic, zhang2024self} employ uniquely crafted prompt templates tailored to individual tasks, along with few-shot learning techniques, to create scripts that can produce the desired response. Thus, for a specific task, the user would need to design a prompt template and provide some learning examples to get the most accurate responses from LLMs.


Figure~\ref{fig:Example} shows an example of a step-by-step execution of \tool. Specifically, given the input prompt \q{Ed had 22 more marbles than Doug. Doug lost 8 of his marbles at the playground. How many more marbles did Ed have than Doug then?}, which asks \tool to perform a subtraction task of \q{Marbles Ed had} and \q{Marbles Doug had after he lost some of them at the playground}. \tool performs step-back analysis to extract the specification of the inputs (i.e., the initial difference in marbles and the amount that Doug had after the accident). At the same time, \tool performs chain-of-thought (CoT) analysis to extract the procedure (i.e., Understanding the situation and the changes after the accident) as well as the expected output (i.e., calculating the new difference). By combining these two pieces of information, \tool was able to generate the solution script that correctly outputs the expected result.

\subsection{Script generation with step-back and zero-shot chain-of-thought prompting}
To reduce the reliance on human effort in utilizing LLMs for task-oriented prompts, \tool leverages two recently proposed prompt engineering techniques: 
step-back prompting~\cite{zheng2023take} and zero-shot chain-of-thought prompting~\cite{kojima2022large}.
Step-back prompting helps \tool analyze the query to identify the relevant inputs and their requirements. Additionally, zero-shot chain-of-thought prompting analyzes the query to extract the relevant steps and procedures to perform the task while requiring no additional input from the user. By integrating these two processes, \tool could produce a script that faithfully represents the original task, which in turn yields a precise response.

\smallskip\noindent\textbf{\textit{Extracting inputs and specification phase}}:
\tool employs step-back prompting to identify the input requirements from the initial prompt. This process involves querying the LLM to identify and outline the specific inputs needed for each task.
By taking a step back, \tool enables the model to focus on the essential inputs
required for successful code execution in the later stage.

Step-back prompting consists of two stages: abstraction and reasoning. Instead of asking the question directly, the abstraction stage asks the LLM a step-back question about a related higher-level concept or principle~\cite{zheng2023take}. In the reasoning stage, the LLM is asked to reason about the high-level concept or principle facts found after the step-back question. In this work, \tool utilizes the abstraction stage to extract inputs and their specifications from the original prompt. It then performs the reasoning stage when generating code based on the extracted inputs.

As demonstrated in Figure~\ref{fig:Example}, \tool performs the abstraction stage by 
asking a step-back question \q{... Take a step back and extract all the inputs mentioned in the client's query ...} to extract the inputs from the original prompt.
Specifically, 
by directing the LLM to focus on a higher concept (i.e., inputs and their specifications)
without directly addressing the original prompt, \tool can extract accurate inputs that help the LLM in the later code generation phase.
In this example, \tool extracts the two inputs (i.e., \q{22} and \q{8}) and their specifications (i.e., \q{initial difference in marbles} and \q{Doug's loss}).
These are then integrated into the generated code later as \code{initial\_difference = 22} and \code{doug\_lost = 8}.

\smallskip\noindent\textbf{\textit{Extracting the goal and procedure steps phase}}:
To improve the precision of the code generation step, \tool applies zero-shot chain-of-thought prompting to 
delve deeper into the goal of the task and the logical steps needed to achieve it while requiring no additional examples. \tool
prompts the LLM to articulate a step-by-step reasoning process, effectively mapping out the pathway from problem statement to solution. As shown in Figure~\ref{fig:Example}, \tool uses the prompt \q{... should guide us step by step on how to solve this problem ...} to extract the procedure steps needed to address the original prompt. By encouraging the LLM to express explicitly the thought process, \tool can extract a detailed goal and the methodological procedure required to complete the task. This process clarifies the objective to ensure the generated code aligns closely with the intended outcome. By utilizing the zero-shot variant of chain-of-thought, \tool eliminates the need for additional examples that prior work such as PAL requires to perform optimally.

Figure~\ref{fig:Example} shows that by instructing the LLM to outline steps toward a goal without directly seeking the final answer, \tool can clarify the main objective \q{Calculate the New Difference}. Additionally, chain-of-thought prompting helps \tool 
uncover logical thinking to solve
the problem
\q{It only changes how many more marbles Ed has compared to Doug}. This phase guides
the LLM to better understand the final goal and process during the code generation phase. This is demonstrated in the final generated code \code{new\_difference = initial\_difference + doug\_lost} which matches the extracted steps.

\smallskip\noindent\textbf{\textit{Code generation phase}}:
Combining the 
information extracted in the two previous phases, \tool 
prompts the LLM
to generate code based on clearly defined inputs, the articulated goal, and well-reasoned steps.
This code generation phase combines the inputs of step-back prompting and the steps from chain-of-though prompting 
to synthesize 
a coherent and functional code output. Specifically, \tool employs the prompt \q{Generate a general Python function to solve the following question for general purpose: \{question\}} to ask LLMs to generate a Python function for the question. Followed up by the prompts \q{This is an example to show you how to think about it and how to break it into smaller steps: ``\{The Output from Goal and Steps Extraction\}''} and \q{ For the inputs, use ``\{The Output from Inputs Extraction\}''} to aggregate the outputs from previous phase into code generation phase.
In this way, \tool is able to guide
the LLM to produce code that is
logically aligned with the task's objectives. For example, \tool obtained the generated script (in Figure~\ref{fig:Example}) given the original prompt \q{Ed had 22 more marbles than Doug. Doug lost 8 of his marbles at the playground. How many more marbles did Ed have than Doug then?}
This example demonstrates the effectiveness of step-back and chain-of-thought prompting in helping LLM generate code with the correct inputs \code{initial\_difference = 22} and \code{doug\_lost = 8}, and precise representation of the task \code{new\_difference = initial\_difference + doug\_lost}.

\smallskip\noindent\textbf{\textit{Post-processing and code execution}}:
Since LLMs return free-form responses, \tool 
employ a rigorous extraction and validation process.
Specifically, \tool utilizes regular expressions to extract the generated code from the responses.
The regular expressions target consistent formatting markers (i.e., ```Python''' for code generated by LLMs to ensure consistent extraction accuracy.
Once the code is extracted, additional post-processing is required such as importing required packages and fixing indentation errors.
The code is then executed automatically and the output is extracted as the final response to the user prompt. To get the result, we first set up rules to identify math questions from the code's output. If the output matches these rules, we take and return this part. If not, we just clean up the output and use that as the final answer. This makes sure we always have a neat and relevant result.
For example, The Final Script in Figure~\ref{fig:Example} is the final generated code that is executed and the output is extracted as 30 (i.e., the value of the variable \code{new\_difference}).

\section{Experimental Setup}

\subsection{Task-oriented datasets}
Drawing from the prior study~\cite{goertzel2023generative}, when faced with a more challenging question on prime numbers that were sufficiently common to have representation on the Internet, the system performed adequately at that time. Hence, we create four task-oriented datasets from scratch to thoroughly evaluate \tool performance.
These include simple task-oriented prompts which we found to be particularly difficult for LLMs to directly address.
This new dataset will benefit future research, given the current trend in prompt construction towards decomposition techniques, such as Least-to-Most Prompting~\cite{zhou2023leasttomost} and Decomposed Prompting~\cite{khot2022decomposed}. Our dataset includes decomposition tasks such as finding, counting, true/false questions, and generative tasks.

Table~\ref{tab:math_dataset} shows the summary of these datasets. Each dataset includes multiple task templates that can be used to generate the prompts and the expected responses. Table~\ref{tab:templates} shows the complete template sets for the task-oriented datasets.

\begin{table*}
    \centering
    \vspace{-5pt}
    \caption{Datasets overview
    }
    \label{tab:math_dataset}
    \vspace{-5pt}
    \resizebox{0.7\linewidth}{!}{
    \begin{tabular}{l r l l}
    \toprule
    Dataset                              & N      & Input                     & Output \\
    \midrule
    GSM8K~\cite{cobbe2021training}       & 1319   & Question                  & Number \\
    GSMHard~\cite{gao2023pal}            & 1319   & Question                  & Number \\
    SVAMP~\cite{patel2021nlp}            & 1000   & Question                  & Number \\
    ASDIV~\cite{miao2021diverse}         & 2096   & Question                  & Number \\
    AddSub~\cite{koncel2016mawps}        & 395    & Question                  & Number \\
    MultiArith~\cite{koncel2016mawps}    & 600    & Question                  & Number \\
    Penguins~\cite{suzgun2022challenging}& 149    & Table + Text + Question   & Number + Text \\

    \midrule
    
    Finding                              & 660    & Question                  & Text   \\
    Counting                             & 1100   & Question                  & Number \\
    True/False                           & 500    & Question                  & Binary \\
    Generative                           & 1100   & Question                  & Text + List \\
    \bottomrule
    \end{tabular}
    }
    \vspace{-5pt}
\end{table*}

\begin{table*}
    \centering
    \caption{The templates to create the four task-oriented datasets}
    \label{tab:templates}
    \resizebox{\linewidth}{!}{
    \begin{tabular}{@{}l p{0.92\linewidth} l@{}}
    \toprule
     Dataset & Prompt Template & Response\\
     \midrule
     
     \multirow{8}{*}{Finding}
     &  Choose the word from the three options provided that does not have ``\{\textbf{word}\}`` within it. The single word given is: ``['\{\textbf{word}\}', '\{\textbf{word}\}', '\{\textbf{word}\}']''.& \multirow{8}{*}{Word}\\
     \cmidrule{2-2}
     &  Taking into account that ``\{\textbf{letter}\}'' is identical to ``\{\textbf{letter}\}``, seek out the word among these three that has the most unique letter count. The words are ``['\{\textbf{word}\}', '\{\textbf{word}\}', '\{\textbf{word}\}']''.& \\
     \cmidrule{2-2}
     &  Assuming ``\{\textbf{word}\}'' has precisely one ``\{\textbf{word}\}'', identify from the list below the word(s) that also contain exactly one ``\{\textbf{word}\}''. The list includes: ``['\{\textbf{word}\}', '\{\textbf{word}\}', '\{\textbf{word}\}']''.& \\
     \cmidrule{2-2}
     &  Among the three words listed, select the one that initiates with ``\textbf{\{letter\}}''. The words for consideration are ``\textbf{\{words\}}''.& \\
     \midrule         
     
     \multirow{8}{*}{Counting}
     &  Excluding words that have fewer than four letters, how many words, spaced apart by 'space', exist in this sentence? The input is: \{\textbf{sentence}\}.& \multirow{8}{*}{Number}\\
     \cmidrule{2-2}
     &  How many numeric characters are found in ``\{\textbf{word}\}''?& \\
     \cmidrule{2-2}
     &  What is the count of ``\{\textbf{letter}\}'' in ``\{\textbf{word}\}'' when ignoring uppercase letters?& \\
     \cmidrule{2-2}
     &  What is the total number of distinct letters in ``\{\textbf{word}\}'', disregarding case?& \\
     \cmidrule{2-2}
     &  How many vowels can be found in the ``\{\textbf{word}\}''& \\
     \midrule
     
     \multirow{11}{*}{True/False}
     & If there is a space in ``\{\textbf{word}\}'', is there any space in ``\{\textbf{word}\}''? If there is a space return ``1'', otherwise return ``0''.&\multirow{9}{*}{0 / 1}\\
     \cmidrule{2-2}
     & Is there a capitalization difference between ``\{\textbf{word}\}'' and ``\{\textbf{word}\}''? If there is a difference return ``1'', otherwise return ``0''.&\\
     \cmidrule{2-2}
     & Does this sentence has more than 3 spaces? ``\{\textbf{sentence}\}'' If there are more than 3 spaces return ``1'', otherwise return ``0''.&\\
     \cmidrule{2-2}
     & Is there any repeated word in the following sentence? ``\{\textbf{sentence}\}'' If there are repeated words return ``1'', otherwise return ``0''.&\\
     \cmidrule{2-2}
     & If we assume the letter ``\{\textbf{letter}\}'' is equal to the letter ``\{\textbf{letter}\}'', is there any spelling difference between ``\{\textbf{word}\}'' and ``\{\textbf{word}\}''? If there is a difference return ``1'', otherwise return ``0''.&\\
     \midrule
     
     \multirow{10}{*}{Generative}
     & Take the first letter of each word within the specified sentence, join these letters to construct and return a new word. Words are spaced apart. The input is: ``\{\textbf{sentence}\}''&\multirow{7}{*}{Word}\\
     \cmidrule{2-2}
     & Switch the initial two letters of the word provided and return the word thus generated. The input is: ``\{\textbf{word}\}''&\\
     \cmidrule{2-2}
     & Replace the final letter of the given word with an 's' and return the newly formed word. The input is: ``\{\textbf{word}\}''&\\
     \cmidrule{2-2}
     & Capitalize the first character of the given word and return the word with the adjustment. The input is: ``\{\textbf{word}\}''&\\
     \cmidrule{2-3}
     & Replace the first letters of the words with each other and return the adjusted versions as the response. The words are: ``['\{\textbf{word}\}', '\{\textbf{word}\}', '\{\textbf{word}\}', ...]''&\multirow{2}{*}{List}\\
     \bottomrule
    \end{tabular}
    
    }
    \vspace{-5pt}
\end{table*}

\smallskip\noindent\textbf{Finding Dataset}:
Given that LLMs have been shown to be strong in natural language tasks,
it was a surprise to see LLMs (including GPT-4), perform poorly on simple Finding tasks where the prompt asks the models to
identify specific patterns and letters within some given text.
This Finding dataset includes
1100 queries that ask for a pattern or a word as a response, we demonstrate prompt templates for generating the Finding dataset in Table~\ref{tab:templates}. 
These types of templates evaluate LLMs' capacity to discern patterns amidst varying textual contexts.

\smallskip\noindent\textbf{Counting Dataset}:
Counting tasks is not a natural fit for LLMs. As demonstrated in the Introduction, GPT models can make mistakes when performing counting tasks. In this dataset, we include 1100 queries that 
require numerical responses. We form the dataset with various prompt templates stated in Table~\ref{tab:templates}.
This dataset is designed to test LLMs' ability to handle counting-related challenges such as enumerating numbers, unique letters, and words.

\smallskip\noindent\textbf{True/False Dataset}:
True/False dataset includes tasks requiring LLMs to
discern the veracity of statements related to natural language processing.
The dataset includes queries that describe the natural language processing tasks on sentences or words. The models are required 
to respond with a binary digit. The dataset consists of 500 samples with auto-generated answers from five prompt templates listed in Table~\ref{tab:templates}. These templates encompass five distinct tasks, focusing on identifying aspects such as spacing, capitalization, repetition of words, and spelling differences within natural language sentences or words. 
Utilizing binary responses reduces the complexity of response composition and focuses the evaluation on the models' ability to analyze and perform natural language tasks.

\smallskip\noindent\textbf{Generative Dataset}:
Performing generative tasks is generally considered a strong point for LLMs which are generative in nature. LLMs are known to generate surprisingly well-structured responses for various creative prompts. However, procedural generative tasks that require strict procedure steps to complete are different from the typical free-form creative tasks that LLMs are known for. 
For example, in a procedural generative task, the LLM could be tasked with creating a word from the first letter of each word in a sentence.
We construct a generative dataset
that aims to assess the capability of models to produce novel responses by following a given procedure and inputs. 
This dataset 
comprises 1100 queries, formatted as prompts which ask
the models to generate a new word as a response in various ways. 
There are five distinct categories
of prompts listed in Table~\ref{tab:templates} that challenge various aspects of textual manipulation, including word swapping, capitalization adjustments, and modifications to the letters at the end of words.
This dataset seeks to thoroughly explore the LLM's generative potential and its ability to understand and manipulate language constructs in novel ways.

\subsection{Mathematical and symbolic reasoning datasets}
Following prior work~\cite{gao2023pal,zhao2023automatic}, we evaluate \tool on 
mathematical and symbolic reasoning datasets such as GSM8K, GSMHard~\cite{cobbe2021training}, SVAMP~\cite{patel2021nlp}, MAWPS~\cite{koncel2016mawps}, PENGUINS~\cite{suzgun2022challenging, gao2023pal}, and ASDIv\cite{miao2021diverse}.

\textbf{GSM8K and GSMHard} are expansive collections of high-quality, linguistically varied grade school mathematics word problems, meticulously curated by adept problem composers. Each problem within the dataset necessitates a solution process that spans between two to eight steps, predominantly involving a series of fundamental arithmetic operations (i.e., addition, subtraction, multiplication, and division) to deduce the conclusive answer.

\textbf{SVAMP} (i.e., Simple Variations on Arithmetic Math Word Problems) dataset presents itself as a challenge set specifically designed for elementary-level Math Word Problems (MWP). An MWP is defined by a concise narrative in Natural Language, delineating a scenario or state of the world, which culminates in posing a query regarding one or more unknown quantities.

\textbf{MAWPS} is an online compendium dedicated to Math Word Problems (MWP) and serves as a comprehensive dataset for the evaluation of diverse algorithms.

\textbf{PENGUINS} delineates a task framework that integrates a tabular dataset of penguins, augmented with supplementary descriptors in natural language. The primary objective within this framework is to deduce answers to queries concerning the attributes of the penguins based on the provided dataset and descriptions.

\textbf{ASDIv} corpus (i.e., Academia Sinica Diverse MWP Dataset) is another MWP dataset which consists
diverse language patterns and problem types, constituting an English MWP collection. This corpus is designed for the assessment of the proficiency of various MWP solvers.


\subsection{Baselines}
\label{sec:baselines}
PAL~\cite{gao2023pal} is the state-of-the-art approach that focuses on task-oriented prompts. PAL employs a code interpreter for problem reasoning with few-shot prompting
while \tool utilizes zero-shot learning. 
To gain a deeper understanding of how our approach compares to the state-of-the-art
we also include PAL's zero-shot prompting technique (PAL ZS) as our second baseline.
PAL ZS utilizes the same prompt template as PAL but
without the use of examples.

Recently, a few variants of PAL have been proposed such as Model Selection~\cite{zhao2023automatic} and X-of-Thoughts~\cite{liu2023plan}. Distinct from \tool, which employs a two-step intermediate process for script creation, 
the Model Selection approach alternates between using CoT or PAL based on which yields more accurate outcomes as determined by language model evaluations. Meanwhile, X-of-Thought adopts the framework consisting of planning and verification phases, selecting the optimal method from among CoT, Program-of-Thought (PoT), and  Equation-of-Thought (EoT) for problem-solving. Since these variants are also using code generation, we adopt their result as our baseline.

Our comparison does not include another variant, Code-based Self-Verification (CSV)~\cite{zhou2023solving}, due to its dependence on the GPT-4 Code Interpreter and its evaluation solely on the MATH dataset. The dataset does not align with the types of problems \tool aims to address.

Follow prior work, 
we also adopt the most recent GPT-4 model (i.e., gpt-4-0125-preview) and a less powerful GPT-3.5 model (i.e., gpt-3.5-turbo-0125) as our backend language model.

\subsection{Experiment Details} \label{experient_detail}
We replicate PAL's result by executing PAL code from their GitHub using the same version of GPT-3.5 and GPT-4 as stated in the original paper. To run PAL on our task-oriented datasets, we form the PAL few-shot prompt with four examples by randomly selecting one example from each task-oriented dataset, otherwise, we keep all PAL's templates the same.

The zero-shot version of PAL uses the same templates but without any examples. Since PAL didn't provide a zero-shot prompt, we developed the zero-shot version of PAL inspired by another related study, PoT~\cite{chen2023program}. We adopt their prompt format by first posing the problem, followed by a request for the LLMs to complete a Python function named ``solution()'' without adding any examples. The function name and the return type are the same as the PAL few-shot version.
For a fair comparison, we utilize the same metric used by PAL, which adopts exact match scores for evaluation.
Unless specified differently, all experiments by default employ greedy decoding, adjusting the temperature of the language models to 0.0.
\section{Result and Discussion}

In this section, we compare \tool against state-of-the-art code generation with zero-shot (RQ1) or few-shot (RQ2).
In RQ3, we evaluate if self-consistency could help improve \tool at a significant cost. Finally, we perform an ablation study (RQ4) to evaluate the contribution of each component in \tool (i.e., input extraction and step extraction).

\subsection{RQ1: How does \tool compare to the state-of-the-art zero-shot prompting approaches with code generation?}

Similar to prior work, \tool employs script generation to solve mathematical and task-oriented problems. Hence, we first assess the effectiveness of \tool when comparing state-of-the-art approaches with code generation in the zero-shot scenario.
Existing code generation methods such as PAL, Model Selection (MS), and X-of-Thought (XoT), all utilize few-shot prompting.
To evaluate how \tool is compared to the prior work in the zero-shot scenario, we create a baseline PAL ZS (PAL zero-shot version) by incorporating a zero-shot prompts template~\cite{chen2023program}
into PAL to be our zero-shot baseline.

\begin{table*}[t]
    \centering
    \caption{
    Comparison of \tool accuracy (\%) to PAL Zero-shot,
    evaluated across eleven benchmarks using two GPT models.
    The best accuracy scores for each dataset and model are 
    bold.
    }
    \label{tab:zeroshot_result}
    \resizebox{0.7\linewidth}{!}{
    \begin{tabular}{l c l@{}c c l@{}c}
        \toprule
         LLM &  \multicolumn{3}{c}{GPT-3.5} &  \multicolumn{3}{c}{GPT-4}\\
         
         \cmidrule(lr){1-1} \cmidrule(lr){2-4} \cmidrule(lr){5-7}
         
         Approach 
         &PAL ZS
         &\multicolumn{2}{c}{\textbf{\tool}}
         &PAL ZS
         &\multicolumn{2}{c}{\textbf{\tool}}\\
         

         \cmidrule(lr){1-1}
         \cmidrule(lr){2-2} \cmidrule(lr){3-4}
         \cmidrule(lr){5-5} \cmidrule(lr){6-7}

         Metric
         & $Acc$
         & \multicolumn{1}{c}{$Acc$} & \multicolumn{1}{c}{$\Delta$}
         & $Acc$ 
         & \multicolumn{1}{c}{$Acc$} & \multicolumn{1}{c}{$\Delta$}\\
         
         \cmidrule(lr){1-1} \cmidrule(lr){2-4} \cmidrule(lr){5-7}
         
         GSM8K &  76.6&\textbf{84.2}&\textcolor{blue}{$\uparrow$7.6}
         & 93.6& \textbf{95.3}&\textcolor{blue}{$\uparrow$1.7}\\
         
         GSMHard&  61.8&\textbf{69.6}&\textcolor{blue}{$\uparrow$7.8}
         &  74.1&\textbf{78.2}&\textcolor{blue}{$\uparrow$4.1}\\
         
         ASDIV&  85.3
         &\textbf{91.4}&\textcolor{blue}{$\uparrow$6.1}
         &  92.7&\textbf{97.2}&\textcolor{blue}{$\uparrow$4.5}\\
         
         SVAMP&  82.8&\textbf{84.3}&\textcolor{blue}{$\uparrow$1.5}
         &  94.0&\textbf{94.8}&\textcolor{blue}{$\uparrow$0.8}\\
         
         AddSub&  \textbf{93.1}&89.8&\textcolor{red}{$\downarrow$3.3}
         &  95.7&\textbf{97.7}&\textcolor{blue}{$\uparrow$2.0}\\
         
         Multiarith&  \textbf{97.3}&96.8&\textcolor{red}{$\downarrow$0.5}
         &96.8&\textbf{98.7}&\textcolor{blue}{$\uparrow$1.9}\\
         
         Penguin&  59.1&\textbf{94.3}&\textcolor{blue}{$\uparrow$35.2}
         &  87.2&\textbf{97.5}&\textcolor{blue}{$\uparrow$10.3}\\

        \cmidrule(lr){1-1} \cmidrule(lr){2-4} \cmidrule(lr){5-7}
         
         Finding& 93.8&\textbf{98.4}&\textcolor{blue}{$\uparrow$4.6}
         & 95.5&\textbf{99.8}&\textcolor{blue}{$\uparrow$4.3}\\
         
         Counting& \textbf{89.1}&87.8&\textcolor{red}{$\downarrow$1.3}
         & 87.5&\textbf{89.8}&\textcolor{blue}{$\uparrow$2.3}\\
         
         True/False& 59.2&\textbf{76.7}&\textcolor{blue}{$\uparrow$17.5}
         & 84.4&\textbf{93.8}&\textcolor{blue}{$\uparrow$9.4}\\
         
         Generative& 84.9&\textbf{94.1}&\textcolor{blue}{$\uparrow$9.2}
         & 98.7&\textbf{99.9}&\textcolor{blue}{$\uparrow$1.2}\\
         
        \cmidrule(lr){1-1} \cmidrule(lr){2-4} \cmidrule(lr){5-7}
        
        Average & 80.3& \textbf{87.9}&\textcolor{blue}{$\uparrow$7.6} 
        & 90.9 
        &\textbf{94.8}&\textcolor{blue}{$\uparrow$3.9}\\
        \bottomrule
    \end{tabular}
    }
    \vspace{-5pt}
\end{table*}

\begin{table*}[t]
    \centering
    \caption{
    Comparison of \tool accuracy (\%) to other Few-shot code generation approaches,
    evaluated across eleven benchmarks using two GPT models.
    The 
    $^*$ symbol 
    indicates the accuracy is from 
    original papers.
    The best accuracy scores for each dataset and model are 
    bold.
    }
    \label{tab:fewshot_result}
    \resizebox{\linewidth}{!}{
    \begin{tabular}{l l@{ }c l@{ }c l@{ }c l@{}c l@{ }c l@{ }c l@{}c}
        \toprule
         LLM &  \multicolumn{8}{c}{GPT-3.5} &  \multicolumn{6}{c}{GPT-4}\\
         
         \cmidrule(lr){1-1} \cmidrule(lr){2-9} \cmidrule(lr){10-15}
         
         Approach  
         &\multicolumn{2}{c}{PAL}
         &\multicolumn{2}{c}{MS$^{*}$}
         &\multicolumn{2}{c}{XoT$^{*}$}
         &\multicolumn{2}{c}{\textbf{\tool}}
         &\multicolumn{2}{c}{PAL}
         &\multicolumn{2}{c}{MS$^{*}$}
         &\multicolumn{2}{c}{\textbf{\tool}}\\
         
         Zero-shot
         &\multicolumn{2}{c}{\ding{55}}
         &\multicolumn{2}{c}{\ding{55}}
         &\multicolumn{2}{c}{\ding{55}}
         &\multicolumn{2}{c}{\ding{51}}
         &\multicolumn{2}{c}{\ding{55}}
         &\multicolumn{2}{c}{\ding{55}}
         &\multicolumn{2}{c}{\ding{51}}\\

         \cmidrule(lr){1-1}
         \cmidrule(lr){2-3} \cmidrule(lr){4-5} \cmidrule(lr){6-7} \cmidrule(lr){8-9}
         \cmidrule(lr){10-11} \cmidrule(lr){12-13} \cmidrule(lr){14-15}

         Metric
         & \multicolumn{1}{c}{$Acc$} & \multicolumn{1}{c}{$T\Delta$}
         & \multicolumn{1}{c}{$Acc$} & \multicolumn{1}{c}{$T\Delta$}
         & \multicolumn{1}{c}{$Acc$} & \multicolumn{1}{c}{$T\Delta$}
         & \multicolumn{1}{c}{$Acc$} & \multicolumn{1}{c}{$A\Delta$}
         & \multicolumn{1}{c}{$Acc$} & \multicolumn{1}{c}{$T\Delta$}
         & \multicolumn{1}{c}{$Acc$} & \multicolumn{1}{c}{$T\Delta$}
         & \multicolumn{1}{c}{$Acc$} & \multicolumn{1}{c}{$A\Delta$}\\
         
         \cmidrule(lr){1-1} \cmidrule(lr){2-9} \cmidrule(lr){10-15}
         
         GSM8K & 81.0&\textcolor{red}{$\downarrow$3.2} & 82.6 &\textcolor{red}{$\downarrow$1.6} & 83.3&\textcolor{red}{$\downarrow$0.9}
         &\textbf{84.2}&\textcolor{blue}{$\uparrow$0.9}
         & 94.8&\textcolor{red}{$\downarrow$0.5}& \textbf{95.6}&\textcolor{blue}{$\uparrow$0.3}
         & 95.3&\textcolor{red}{$\downarrow$0.3}\\
         
         GSMHard& 64.1& \textcolor{red}{$\downarrow$5.5}&  -& &  63.4& \textcolor{red}{$\downarrow$6.2}
         &\textbf{69.6}&\textcolor{blue}{$\uparrow$5.5}
         &  70.9& \textcolor{red}{$\downarrow$7.3}&  -& 
         &\textbf{78.2}&\textcolor{blue}{$\uparrow$4.1}\\
         
         ASDIV&  86.4& \textcolor{red}{$\downarrow$5.0}&  89.4& \textcolor{red}{$\downarrow$2.0}&  -&
         &\textbf{91.4}&\textcolor{blue}{$\uparrow$2.0}
         &  92.1& \textcolor{red}{$\downarrow$5.1}&  93.5& \textcolor{red}{$\downarrow$3.7}
         &\textbf{97.2}&\textcolor{blue}{$\uparrow$3.7}\\
         
         SVAMP&  \textbf{84.6}& \textcolor{blue}{$\uparrow$0.3}&  84.3&          \textcolor{gray}{-0.0}&  83.6& \textcolor{red}{$\downarrow$0.7}
         &84.3&\textcolor{red}{$\downarrow$0.3}
         &  94.6& \textcolor{red}{$\downarrow$0.2}&  93.7& \textcolor{red}{$\downarrow$1.1}
         &\textbf{94.8}&\textcolor{blue}{$\uparrow$0.2}\\
         
         AddSub&  \textbf{92.7}& \textcolor{blue}{$\uparrow$2.9}&  90.6&          \textcolor{blue}{$\uparrow$0.8}&  90.5& \textcolor{blue}{$\uparrow$0.7}
         &89.8&\textcolor{red}{$\downarrow$3.3}
         &  97.2& \textcolor{red}{$\downarrow$0.5}&  95.7& \textcolor{red}{$\downarrow$2.0}
         &\textbf{97.7}&\textcolor{blue}{$\uparrow$0.5}\\
         
         Multiarith&  \textbf{98.7}& \textcolor{blue}{$\uparrow$1.9}&  98.7&          \textcolor{blue}{$\uparrow$1.9}& 97.3&\textcolor{blue}{$\uparrow$0.5}
         &96.8&\textcolor{red}{$\downarrow$1.9}
         &  98.8& \textcolor{blue}{$\uparrow$0.1}&  \textbf{99.0}& \textcolor{blue}{$\uparrow$0.3}
         &98.7&\textcolor{red}{$\downarrow$0.3}\\
         
         Penguin&\textbf{96.6}&\textcolor{blue}{$\uparrow$2.3}&  -&   &  -&         
         &94.3&\textcolor{red}{$\downarrow$2.3}
         &  96.6&          \textcolor{red}{$\downarrow$0.9}&  -&         
         &\textbf{97.5}&\textcolor{blue}{$\uparrow$0.9}\\

        \cmidrule(lr){1-1} \cmidrule(lr){2-9} \cmidrule(lr){10-15}
         
         Finding& 97.7& \textcolor{red}{$\downarrow$0.7}& -&& -&
         &\textbf{98.4}&\textcolor{blue}{$\uparrow$0.7}
         & \textbf{99.8}&         \textcolor{gray}{-0.0}& -&
         &\textbf{99.8}&\textcolor{gray}{-0.0}\\
         
         Counting& 84.6& \textcolor{red}{$\downarrow$3.2}& -&& -&
         &\textbf{87.8}&\textcolor{blue}{$\uparrow$3.2}
         & 88.5& \textcolor{red}{$\downarrow$1.3}& -&
         &\textbf{89.8}&\textcolor{blue}{$\uparrow$1.3}\\
         
         True/False& 76.0& \textcolor{red}{$\downarrow$0.7}& -&& -&
         &\textbf{76.7}&\textcolor{blue}{$\uparrow$0.7}
         & \textbf{95.6}& \textcolor{blue}{$\uparrow$1.8}& -&
         &93.8&\textcolor{red}{$\downarrow$1.8}\\
         
         Generative& 87.7& \textcolor{red}{$\downarrow$6.4}& -&& -&
         &\textbf{94.1}&\textcolor{blue}{$\uparrow$6.4}
         & 99.4& \textcolor{red}{$\downarrow$0.5}& -&
         &\textbf{99.9}&\textcolor{blue}{$\uparrow$0.5}\\
         
        \cmidrule(lr){1-1} \cmidrule(lr){2-9} \cmidrule(lr){10-15}
        
        Average  
        & 86.4&\textcolor{red}{$\downarrow$1.5}
        & -& 
        & -& 
        & \textbf{87.9}&\textcolor{blue}{$\uparrow$1.5} 
        
        & 93.5&\textcolor{red}{$\downarrow$1.3}& -&
        &\textbf{94.8}&\textcolor{blue}{$\uparrow$1.3}\\
        \bottomrule
    \end{tabular}
    }
    \vspace{-5pt}
\end{table*}

Table~\ref{tab:zeroshot_result} compares \tool's accuracy to other state-of-the-art zero-shot approaches (Row Approach) with code generation across 11 datasets (Columns \textit{Acc}). The table is divided into two sections each corresponding to a version of GPT (e.g., GPT-3.5 and GPT-4).
It includes the $\Delta$ column to show the relative performance differences between \tool and PAL ZS. For instance, \tool paired with GPT-4 outperforms the state-of-the-art zero-shot approach PAL ZS in all 11 datasets with a margin as big as 10.3\% on the Penguin dataset. When a weaker LLM model such as GPT-3.5 is used, \tool still outshines PAL ZS on 8 over 11 datasets with margins as big as 35.2\% in the case of the Penguin dataset. The reason weaker LLM models fall short on some datasets is that the less advanced GPT model is unable to leverage the additional stages that TITAN offers. In scenarios involving simple datasets, inundating simpler models with excessive information can lead to incorrect responses.~\cite{chiang2024over}

The average row specifies the average performance of each approach across 11 datasets. 
On average, \tool achieved state-of-the-art zero-shot performance with significant margins of 7.6\% and 3.9\% when used on GPT-3.5 and GPT-4 respectively. Overall, the weaker LLM models such as GPT-3.5 benefit more from the additional stages that \tool utilizes. In contrast, the stronger models such as GPT-4 have better reasoning capability build-in and hence benefit less from step-back and chain-of-thought prompting. This result highlights \tool's ability to significantly boost the performance of weaker language models.


When looking at the GPT-4 result, \tool improvement on PAL ZS is relatively small (around 2\%) on simpler datasets with small numbers and easy questions such as AddSub, Multiarith, and Counting. On the other hand, on more challenging datasets involving large numbers, tables, and complicated questions such as GSMHard, True/False, and Penguin datasets, \tool significantly outperforms PAL-ZS (by 4.1\%, 9.4\%, and 10.3\% respectively)
This suggests that the \tool is particularly suitable for complex task-oriented prompts.

\begin{tcolorbox}[boxrule=0.5pt, colback=gray!10, arc=4pt,left=6pt,right=6pt,top=6pt,bottom=6pt,boxsep=0pt]
\textbf{Finding 1:} \tool outperforms the state-of-the-art zero-shot approach by 7.6\% and 3.9\% when paired with GPT-3.5 and GPT-4. In some cases, the performance gain can be as big as 35.2\% and 10.3\% with GPT-3.5 and GPT-4 respectively. 
\end{tcolorbox}

\subsection{RQ2: How does \tool compare to the state-of-the-art few-shot prompting approaches with code generation?}
In this RQ, we assess the broader effectiveness of \tool when comparing state-of-the-art approaches with code generation that also utilizes few-shot prompting such as PAL few-shot version, Model Selection (MS), and X-of-Thought (XoT).
The details of each baseline are discussed in Sec~\ref{sec:baselines}. We replicate PAL on the original and our task-oriented datasets, we employ the code provided in the original study.

Table~\ref{tab:fewshot_result} compares \tool's accuracy to other state-of-the-art Few-shot approaches with code generation.
The asterisk \q{*} symbol indicates approaches that are not possible to replicate (i.e., due to missing executable source code) and for which the accuracy values are taken from the original papers. This is why the Table does not show results for MS and XoT on our proposed task-oriented datasets which is indicated by the dash (\q{-}).

In Table~\ref{tab:fewshot_result}, for each combination of LLM and dataset, the best accuracy value is highlighted in bold, indicating state-of-the-art performance. The $A\Delta$ columns illustrate the improvement that \tool makes when compared to all baselines. For example, \tool achieves state-of-the-art performance on 8 over 11 datasets when paired with GPT-4 LLM while only marginally losing to few-shot approaches (which needed additional non-trivial examples) on three datasets GSM8K, Multiarith, and True/False (by 0.3\%, 0.3\% and 1.8\% respectively as indicated by the $A\Delta$ column).

Overall, when paired with the most advanced LLM (i.e., GPT-4), \tool provides the most enhancement among all approaches across most datasets. This can be seen by looking at the $T\Delta$ columns which indicate the difference between each baseline and \tool. Only in four cases where PAL and MS (which requires non-trivial examples) outperform \tool with marginal gaps (i.e., 0.1\%, 0.3\%, 0.3\%, and 1.8\%).

 On average, \tool with GPT-4 achieves an average performance boost of 1.3\% compared to PAL (which unlike \tool requires few-shot examples). When evaluated on GPT-3.5, \tool performs better on average than alternative few-shot approaches, achieving an average accuracy enhancement of 1.5\% over PAL. These results indicate that \tool achieves state-of-the-art performance in most cases even against few-shot approaches that require non-trivial examples.

When moving from GPT-3.5 to GPT-4, the average performance gain \tool provides over PAL is only 0.2\%(between 1.5\% and 1.3\%) which is significantly smaller than the 3.7\% (between 7.6\% and 3.9\%) gain \tool provides over PAL ZS.
This might be because the more advanced language models (i.g., GPT-4) benefit more from few-shot prompting and hence are harder to enhance with step-back and chain-of-thought prompting that \tool employs.
Nonetheless, \tool enhances the performance of these models across the board. 

\begin{tcolorbox}[boxrule=0.5pt, colback=gray!10, arc=4pt,left=6pt,right=6pt,top=6pt,bottom=6pt,boxsep=0pt]
\textbf{Finding 2:} Overall, \tool achieves state-of-the-art performance in most (8 out of 11) cases when paired with GPT-4 while only marginally losing to few-shot approaches (which needed additional non-trivial examples) on three datasets by small margins. Overall, \tool outperforms PAL by 1.3\%.
\end{tcolorbox}

To better demonstrate how \tool was able to effectively generate correct responses and illustrate \tool's limitations we include two examples where: \textbf{1:} \tool is correct while PAL fails and \textbf{2:} both PAL and \tool fail.

\begin{figure}
    \centering
    \includegraphics[width=0.9\linewidth]{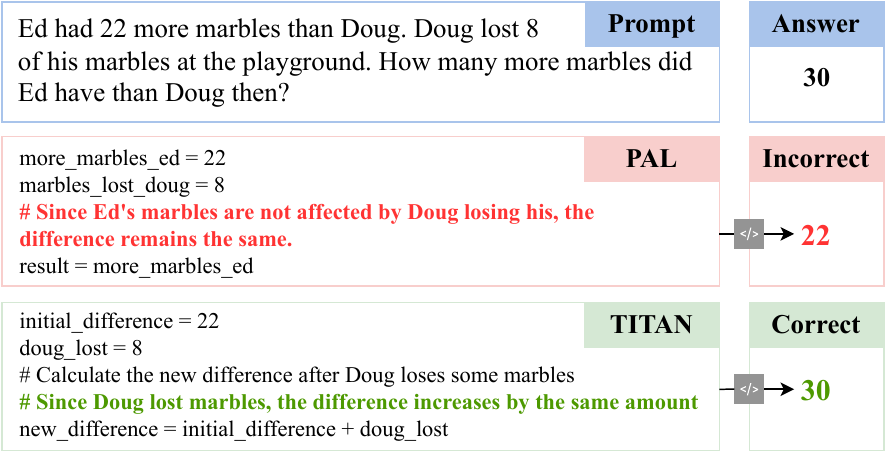}
    \vspace{-7pt}
    \caption{An example prompt 
    where \tool successes
    while PAL fails.}
    \label{fig:discussion}
    \vspace{-10pt}
\end{figure}

\smallskip\noindent\textbf{Case 1}:
Figure~\ref{fig:discussion} shows an example from the ASDIV dataset where \tool can combine correct reasoning from the input and step extraction phase to generate the correct code. Specifically, \tool is able to identify the inputs \code{initial\_difference} and \code{doug\_lost} as well as the importance of reasoning: \code{\#Since Doug lost marbles, the difference increases by the same amount} which is realized by the important extracted step: \code{new\_difference = initial\_difference + doug\_lost}.
This is further illustrated by Figure~\ref{fig:Example} where the extraction of goals and steps yields the correct logical steps: \q{Calculate the New Difference}
On the other hand, without either of these reasoning phases, PAL fails to generate the correct responses.
Specifically, PAL identifies the input name wrong \code{more\_marbles\_ed} which hinders the integration reasoning into the solution. This example showcases the distinct differences between PAL and \tool and highlights \tool's profound grasp of the problem context due to the reasoning phases.

\begin{figure}
    \centering
    \includegraphics[width=0.9\linewidth]{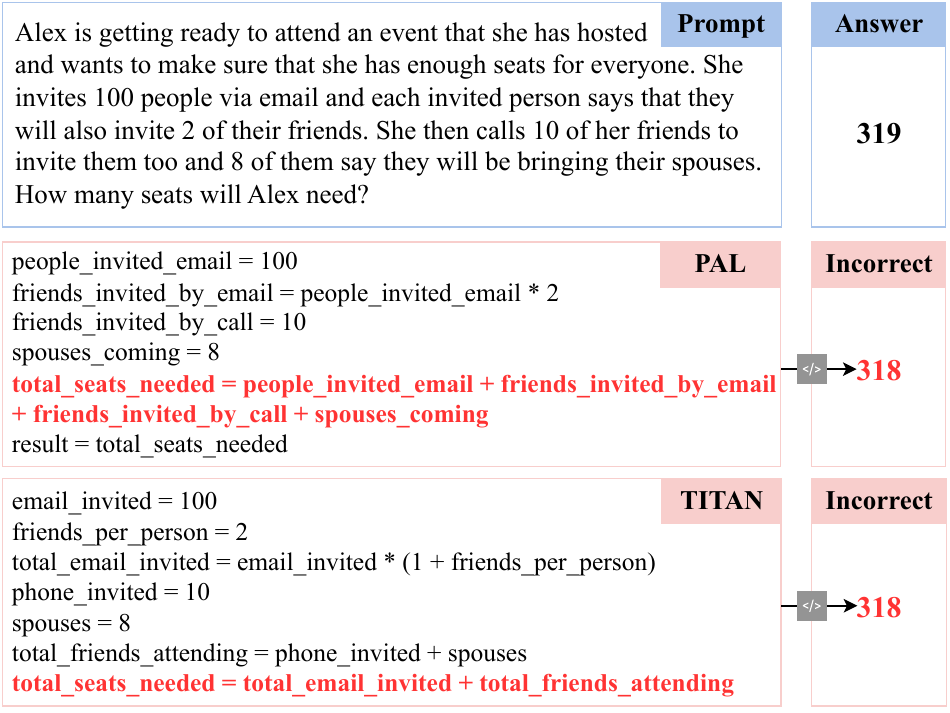}
    \vspace{-7pt}
    \caption{An example prompt 
    where both PAL and \tool fail.}
    \label{fig:discussion2}
    \vspace{-2pt}
\end{figure}

\smallskip\noindent\textbf{Case 2}:
Figure~\ref{fig:discussion2} shows an example where both PAL and \tool fail to generate the correct answer. Both methods correctly reason about the composition of the answer by counting the email invitations, the email invitations' friends, the phone invitations, and their spouses. However, both fail to account for one additional seat for Alex herself. This example might expose \tool's inherent weakness from the underlying GPT model (i.e., GPT-4) where the GPT model does not link the number of seats needed to the number of attendants and instead links to the number of invitees mentioned in the prompt.
We suspect that this can be due to GPT's autoregressive mechanism (i.e., tokens are generated sequentially) that makes each token's production reliant heavily on the outcomes generated before it~\cite{xu2023compress}.
This highlights unique challenges and flaws in the way GPT-4 learns and encodes these processes~\cite{goertzel2023generative}.

\subsection{RQ3: Can self-consistency help improve \tool?}
Self-consistency~\cite{wang2023selfconsistency} is an agnostic strategy that can be applied to CoT prompting approaches to improve the underlying approach performance. One downsize of self-consistency is that it requires many duplicated queries to be sent to the LLM which can have diminishing returns~\cite{huang2023large}.
Since \tool incorporates the CoT we hypothesize that incorporating self-consistency could further improve \tool's performance. However, due to the elevated cost, we do not incorporate self-consistency in \tool by default and instead use the temperature of 0.

\begin{table}
    \centering
    \vspace{-5pt}
    \caption{\tool self-consistency integration on 
    GSM8K, Multiarith, and  True/False datasets utilizing GPT-4 with
    majority voting from three samples.
    }
    \label{tab:self-consistency}
    \vspace{-5pt}
    \begin{tabular}{l c c}

    \toprule
    
    & \tool & \tool + SC@3\\


    
    \midrule
    
    GSM8K& 95.3 & 95.6 \textcolor{blue}{$\uparrow$0.3}\\
    Multiarith& 98.7&99.5 \textcolor{blue}{$\uparrow$0.8}\\
    True/False& 93.8&95.4 \textcolor{blue}{$\uparrow$1.6}\\
    \bottomrule

    \end{tabular}
    

\end{table}

In this RQ3, we want to test if self-consistency can help improve \tool further and if the benefit warrants the cost.
Specifically, we apply self-consistency on \tool for three datasets (GSM8K, Multiarith, and True/False) to see if self-consistency can help \tool achieve state-of-the-art performance. As indicated by the original paper~\cite{wang2023selfconsistency}, we set the LLM's temperature to a nonzero value to retrieve distinct response samples. In this case, we set the GPT-4 temperature 
to 0.7 before running \tool three times.
The final response is subsequently chosen by majority voting.


In Table~\ref{tab:self-consistency}, \tool + SC\@3 (self-consistency with three samples) indicates the accuracy (\%) of \tool when self-consistency is integrated with three responses. When comparing the vanilla \tool, self-consistency helps enhance \tool by 0.3\% on GSM8K, 0.8\% on Multiarith, and 1.6\% on True/False at the cost of triple the number of queries. If the cost of self-consistency is not a factor, \tool + SC\@3 would achieve state-of-the-art performance on GSM8K and Multiarith while getting within 0.2\% of the best approach on True/False. 
Overall, the result suggests the adoption of self-consistency not only bolsters \tool's robustness but also achieves superior performance relative to greedy decoding (i.e., setting the LLM temperature to 0.0 which is \tool's default setting) at a significant cost.


\begin{tcolorbox}[boxrule=0.5pt, colback=gray!10, arc=4pt,left=6pt,right=6pt,top=6pt,bottom=6pt,boxsep=0pt]
\textbf{Finding 3:} Overall, Self-consistency can improve \tool performance with significant cost. Specifically, with triple the cost, \tool + SC@3 is shown to improve \tool by achieving state-of-the-art performance on GSM8K and Multiarith.
\end{tcolorbox}

\begin{table}
    \centering
    \vspace{-5pt}
    \caption{\tool Ablation Study
    }
    \label{tab:ablation-study}
    \vspace{-5pt}
    \begin{tabular}{l r@{ }l r@{ }l c}
    \toprule
    Dataset 
    & \multicolumn{2}{c}{\makecell{W/o Input\\Extraction}}  
    & \multicolumn{2}{c}{\makecell{W/o Step\\Extraction}}  
    & \tool\\
    \midrule
    GSM8K&95.6&\textcolor{blue}{$\uparrow$0.3}&93.4&\textcolor{red}{$\downarrow$1.9}&95.3\\
    GSMHard &77.5&\textcolor{red}{$\downarrow$0.7}& 76.1&\textcolor{red}{$\downarrow$2.1}&78.2\\
    ASDIV&95.8&\textcolor{red}{$\downarrow$1.4}& 95.3&\textcolor{red}{$\downarrow$1.9}&97.2\\
    SVAMP&94.5&\textcolor{red}{$\downarrow$0.3}&92.9&\textcolor{red}{$\downarrow$1.9}
    &94.8\\
    AddSub&97.7&\textcolor{gray}{-0.0}&97.7&\textcolor{gray}{-0.0}
    &97.7\\
    Multiarith&98.5&\textcolor{red}{$\downarrow$0.2}&98.3&\textcolor{red}{$\downarrow$0.5} 
    &98.7\\
    Penguins&95.3&\textcolor{red}{$\downarrow$2.2}&94.6&\textcolor{red}{$\downarrow$2.9} 
    &97.5\\
    \midrule
    Finding&98.5&\textcolor{red}{$\downarrow$1.5}&98.3&\textcolor{red}{$\downarrow$1.3} 
    &99.8\\
    Counting&93.6&\textcolor{blue}{$\uparrow$3.8}&88.6&\textcolor{red}{$\downarrow$1.2}
    &89.8\\
    True/False&94.5&\textcolor{blue}{$\uparrow$0.7}&95.1&\textcolor{red}{$\uparrow$1.3}
    &93.8\\
    Generative&99.0&\textcolor{red}{$\downarrow$0.9}&99.1&\textcolor{red}{$\downarrow$0.8}
    &99.9\\

    \midrule

    Average & 94.5&\textcolor{red}{$\downarrow$0.3}& 93.5&\textcolor{red}{$\downarrow$1.4}& 94.8\\
    
    \bottomrule
    \end{tabular}
    \vspace{-5pt}
\end{table}

\subsection{RQ4: How each component contribute to \tool's overall performance?}
Input and step extraction are two important phases that help \tool generate precise scripts that can be used to generate correct responses. In this RQ, we perform an ablation study with GPT-4 to see how each component contributes to \tool's overall performance.

Table~\ref{tab:ablation-study} shows how turning off one of the main reasoning components (Column W/o Input Extraction and W/o Step Extraction) can reduce the effectiveness of \tool. The $\downarrow$ and $\uparrow$ indicate that removing the component reduces or increases the effectiveness of \tool respectively. For example, removing the step extraction phase reduces \tool accuracy on the Penguins dataset by 2.9\%. On all datasets, removing the step extraction phase reduces \tool's performance significantly with an average reduction in performance of 1.5\%. The result suggests that step extraction significantly contributes to \tool's performance.

\begin{figure}  
    \centering
    \includegraphics[width=0.9\linewidth]{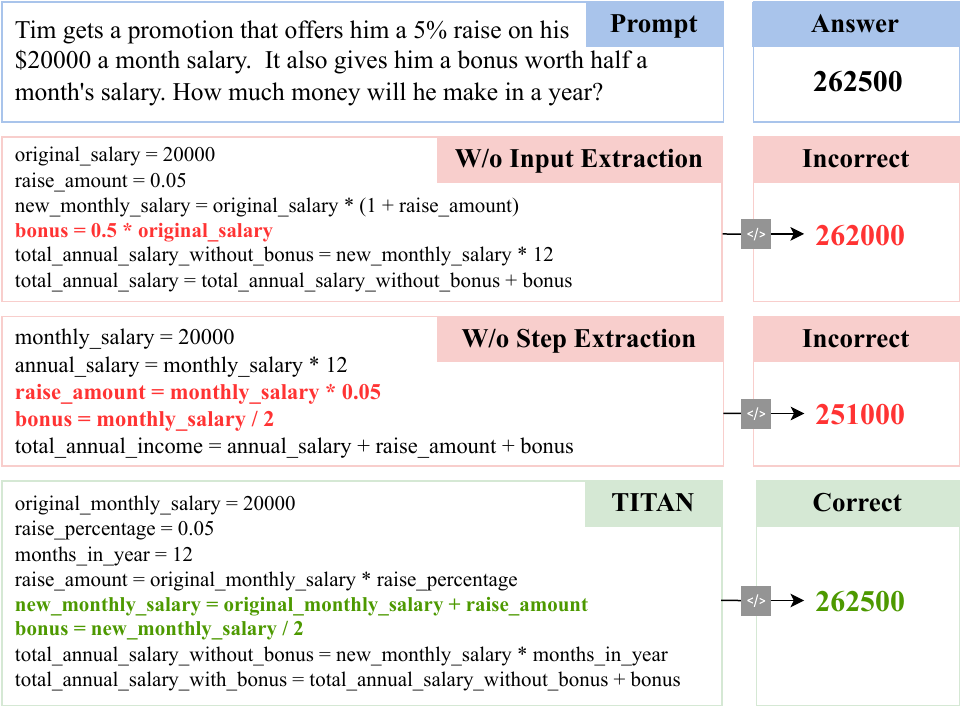}
    \vspace{-7pt}
    \caption{An example 
    where both reasoning phases together help \tool generate the correct response.}
    \label{fig:discussion3}
    \vspace{-10pt}
\end{figure}

On the other hand, removing the input extraction phase hinders \tool's performance in a few cases. For example, on the Counting dataset, \tool gains 3.8\% without the input extraction phase. This could be because, for the counting tasks, there are fewer inputs and including additional information about the input complicates the script generation prompt and affects the final script. This result suggests that input extraction might have less contribution to the overall success of \tool in correctly crafting the answers. However, in a broader term, input extraction is still a necessary component of \tool as removing it reduces \tool performance by an average of 0.3\%.

On AddSub, removing either component does not trigger significant accuracy changes (i.e., changes larger than 0.1\%). This suggests that each reasoning phase already provides enough information to enhance the code generation. This is evident in Table~\ref{tab:fewshot_result}, where \tool outperforms PAL (without such reasonings) by 0.5\% on the AddSub dataset.
Overall, both components are essential to the overall performance of \tool as removing either reduces the overall performance of \tool.




This is further illustrated by the example in Figure~\ref{fig:discussion3}, where without either input and step extraction, \tool failed to utilize the correct input or follow the precise procedure. Specifically, without input extraction, the framework fails to distinguish between the \code{original\_salary} and the \code{new\_monthly\_salary}, leading to incorrect bonus calculations.
On the other hand, without step extraction, the model fails to recognize the correct steps to compute the \code{raise\_amount} (i.e., should be across 12 months) and \code{bonus} (should be on raised salary).
When utilizing both reasoning phases, \tool was able to correctly extract precise inputs and procedural steps to complete the task in a clear and precise manner.


\begin{tcolorbox}[boxrule=0.5pt, colback=gray!10, arc=4pt,left=6pt,right=6pt,top=6pt,bottom=6pt,boxsep=0pt]
\textbf{Finding 4:} Step extraction phase plays a significant role in enhancing \tool's performance while in some cases input extraction can hinder the performance of the framework. However, both components are essential to the overall performance of \tool as removing either reduces the overall performance of \tool.
\end{tcolorbox}

\section{Related Work}

\subsection{Prompt engineering with code generation}
There has been prior work that utilizes code generation to address the gap in LLMs' execution ability when it comes to task-oriented prompts. PAL\cite{gao2023pal} introduces an innovative method for tackling mathematical problems through script generation combined with few-shot prompting. Further improvement is made by a Model Selection technique (MS)~\cite{zhao2023automatic} which employs both PAL and CoT (Chain-of-Thought) in tandem by
selecting the best response between them.

X-of-Thoughts (XoT)~\cite{liu2023plan} represents another code generation strategy, focusing on resolving mathematical and algebraic equations by dynamically switching among different prompting methods.
Another distinct method, CSV~\cite{zhou2023solving}, approaches the resolution of mathematical challenges through coding, which heavily relies on the GPT-4 Code Interpreter. Its efficacy is evaluated exclusively on the MATH dataset~\cite{hendrycks2021measuring}, indicating a specialized focus on coding solutions for mathematical issues.

In contrast, \tool diverges significantly from these methods by adopting a zero-shot learning technique, which enhances its capacity for generalization across diverse problem sets. 
\tool aims to solve reasoning questions without relying on any hand-crafted data, few-shot learning techniques, or human annotators.

\subsection{Traditional prompt engineering without code generation}
The exploration of concepts such as Chain-of-Thought~\cite{wei2022chain} and Step-Back~\cite{zheng2023take} has revealed the capacity of LLMs for zero-shot learning~\cite{kojima2022large} primarily through their ability to process information in a step-by-step manner.
Prior work such as PHP~\cite{zheng2023progressive}, Self-Contrast~\cite{zhang2024self}, and Boosting-of-Thought (BoT)~\cite{chen2024boosting} has been developed to uncover the reasoning paths through post-hoc strategies of prompt engineering, aiming to enhance the models' problem-solving capabilities by mimicking human-like~\cite{he2024exploring} reasoning processes. These techniques use self-verification methods. For example, BoT iteratively generates, assesses, and refines thoughts using the model's self-evaluation. Self-Contrast exploits diverse solutions to enrich reasoning, while PHP refines reasoning paths with LLM outputs toward the correct answer.

Because the mechanisms and effectiveness of self-correction in LLMs are not well-understood~\cite{huang2023large}, we do not include such iterative mechanism in \tool.
Recent research indicates that LLMs are not yet capable of self-correction~\cite{huang2023large}. Distinct from such interactive methods which require human annotations or many more query rounds, 
\tool does not require manual effort to incorporate evaluation information into the reasoning process.

\subsection{Code generation with LLMs}
Recently, research in using LLMs for code generation~\cite{roziere2023code, chen2021evaluating, fried2022incoder, yang2023deep} has made significant advancements. These advancements often come from training these models with more specific examples of code, which makes them better at specific coding tasks~\cite{chen2021evaluating}.
Another line of research is guided code generation~\cite{zhao2022gap, zheng2023outline} where LLMs are utilized to efficiently create code. Specifically, Willard et al.~\cite{willard2023efficient} introduce a system that guides LLMs to produce text using rules and structures from programming languages, reducing unnecessary steps in creating code sequences. Another recent work, SynCode~\cite{ugare2024improving}, is a framework that improves how LLMs understand and generate code by focusing on the rules of programming languages. This approach helps create more accurate code by filtering out mistakes and focusing on the correct coding syntax. \tool differs from these code generation research in the general nature of \tool's input (i.e., prompt) where \tool is designed to improve the overall question/answer capability of LLMs by utilizing script generation and not to solve general software engineering problems related to code generation.

\section{Conclusion}

In this work, we introduced \tool, a novel approach for natural language reasoning that leverages script generation through the extraction of inputs and steps by utilizing Step-Back and Chain-of-Thought prompting respectively. We evaluate \tool on 11 datasets with a comprehensive comparison with prior state-of-the-art. Unlike preceding approaches that predominantly rely on few-shot prompting techniques, \tool employs a zero-shot prompting strategy, thereby eliminating the requirement for hand-crafted data. Our findings demonstrate that \tool exhibits superior performance in a diverse set of tasks. Furthermore, the integration of \tool with self-consistency 
further enhances its efficacy (with some additional cost), thereby underscoring the potential of \tool as a robust solution for advanced natural language reasoning challenges.

\section{Data Availability}
We release our code and data through the following link: \url{https://anonymous.4open.science/r/TITAN-Task-oriented-Prompt-Improvement-with-Script-Generation-3BE4}.

\newpage
\bibliographystyle{ACM-Reference-Format}
\bibliography{refs}
\end{document}